\documentclass[aps,twocolumn,showpacs,floatfix]{revtex4}

\usepackage{graphicx,amsmath}

\newcommand{\eq}[1]{Eq.(\ref{#1})}
\newcommand{\fig}[1]{Fig.~\ref{#1}}

\newcommand{\avg}[1]{\langle #1 \rangle}
\newcommand{\olcite}[1]{Ref.~\onlinecite{#1}}
\newcommand{\ahum}[1]{``#1''}

\newcommand{\rs}{R_{\rm s}}

\newcommand{\COMMENT}[1]{ }

\begin{document}

\title{Nematics with quenched disorder : violation of self-averaging}

\author{J. M. Fish and R. L. C. Vink}

\affiliation{Institute of Theoretical Physics, Georg-August-Universit\"at 
G\"ottingen, Friedrich-Hund-Platz~1, 37077 G\"ottingen, Germany}

\date{\today}

\begin{abstract} We consider the isotropic-to-nematic transition in liquid 
crystals confined to aerogel hosts, and assume that the aerogel acts as a random 
field. We generally find that self-averaging is violated. For a bulk transition 
that is weakly first-order, the violation of self-averaging is so severe, even 
the correlation length becomes non-self-averaging: no phase transition remains 
in this case. For a bulk transition that is more strongly first-order, the 
violation of self-averaging is milder, and a phase transition is observed. 
\end{abstract}

%% 05.70.Fh Phase transitions: general studies
%% 75.10.Hk Classical spin models
%% 64.60.-i General studies of phase transitions
%% 64.70.mf Theory and modeling of specific liquid crystal transitions, including computer simulation 
%% 61.30.Jf Defects in liquid crystals 
%% 61.30.Cz Molecular and microscopic models and theories of liquid crystal structure 
%% 64.70.mf Theory and modeling of specific liquid crystal transitions
%% 61.20.Gy Theory and models of liquid structure
%% 68.43.Mn Adsorption kinetics 

%% 61.30.Pq Microconfined liquid crystals including droplets, cylinders, 
%% randomly confined liquid crystals, polymer dispersed liquid crystals and porous systems 


\maketitle

Liquid crystals confined to quenched disordered media are frequently encountered 
and of practical importance~\cite{citeulike:6711617}. In certain cases -- the 
prototype example being silica aerogel -- the disordered medium induces a 
quenched random field \cite{citeulike:5091293, citeulike:6721301, 
citeulike:5930299, citeulike:6642489, citeulike:6645610, citeulike:6642471, 
citeulike:6642981}. The random field couples to the liquid crystal at 
(essentially) arbitrary locations, and imposes a preferred orientation of the 
nematic director at these locations. One consequence of random field disorder in 
liquid crystals is the loss of long-range nematic order in all experimentally 
relevant dimensions $d \leq 3$ \cite{imry.ma:1975, citeulike:5717604, 
citeulike:6645961, citeulike:6657736, citeulike:5967865}. This, however, does 
not rule out the existence of phase transitions. In contrast, the latter are 
routinely observed \cite{citeulike:5930299, citeulike:5091293, 
citeulike:6721317, citeulike:6642981} and understanding the influence of random 
field disorder on liquid crystal phase transitions is an important topic. One 
known effect is that random fields can change the order of a transition 
\cite{citeulike:4197313, citeulike:6721301}. The bulk isotropic-to-nematic (IN) 
transition in three dimensions (3D) is usually first-order, but random fields 
can render this transition continuous \cite{citeulike:5091293, 
citeulike:6642433, citeulike:6645630, citeulike:5930299} or wipe it out 
completely \cite{citeulike:6721301}. Other known effects include slow dynamics 
\cite{citeulike:6657729, citeulike:6642390, citeulike:6642489, rotunno:097802}, 
lowering of phase transition temperatures \cite{citeulike:6721301, 
citeulike:6642489}, and formation of multidomain nematic structures 
\cite{citeulike:5930299, citeulike:5887661}.

It is also known that systems exposed to random fields generally do not 
self-average: results obtained for one sample of disorder, even if the sample is 
large, are not necessarily representative for all disorder samples 
\cite{aharony.harris:1996, citeulike:6672115, citeulike:6348195}. To what extent 
lack of self-averaging plays a role at the IN transition is the topic of the 
present Letter. Our main result is that, for a bulk IN transition in 3D that is 
weakly first-order, i.e.~the experimentally most relevant case, the violation of 
self-averaging in the presence of random fields is so severe, even the 
correlation length becomes a non-self-averaging quantity 
\cite{citeulike:7465852, citeulike:3201143}. The IN transition temperature, as 
characterized by the temperature of the specific heat maximum, does not become 
sharp in the thermodynamic limit, but is given by a distribution of finite 
width. Hence, no sharp phase transition remains.

To illustrate this point, we have simulated the sprinkled silica spin (SSS) 
model \cite{citeulike:6711320}; models such as this are routinely used to 
describe nematics in disordered media \cite{citeulike:6645961, 
citeulike:5091220, citeulike:5091176, citeulike:5717604, citeulike:6711320}. The 
SSS model is defined on a 3D periodic $V = L \times L \times L$ lattice. A 3D 
unit vector $\vec{d}_i$ (spin) is attached to each lattice site $i$. The energy 
density is given by
\begin{equation}\label{ll}
 \epsilon = -J/V \sum_{\avg{i,j}} | \vec{d}_i \cdot \vec{d}_j|^p, 
 \quad J>0,
\end{equation}
with the sum over nearest neighbors (in what follows, the temperature $T$ is 
expressed in units of $J/k_B$, with $k_B$ the Boltzmann constant). We set $p=2$ 
for now; \eq{ll} then resembles the Lebwohl-Lasher model \cite{physreva.6.426}, 
which undergoes a weak first-order IN transition from a high-$T$ isotropic phase 
(exponential decay of the nematic correlation function to zero), to a low-$T$ 
nematic phase with long-range order (exponential decay of the nematic 
correlation function to a finite positive value). In the SSS model, quenched 
disorder is introduced by marking a fraction $q$ of randomly selected spins as 
quenched (we use $q=0.1$ always). These spins are oriented randomly at the start 
of the simulation and remain static thereafter, which can be conceived as a 
random field of infinite strength acting on a fraction of the spins. Even though 
the random field strength is infinite, $q=0.1$ remains in the weak field limit, 
in the sense that the non-quenched spins still form a percolating network. If 
$q$ is set above the percolation threshold, any phase transition gets trivially 
blocked, since then the correlations cannot propagate through the lattice 
anymore. The SSS model is different from the random-field Ising model because 
the spins are 3D continuous vectors, as opposed to discrete integers. The SSS 
model does not support long-range nematic order at any finite 
temperature~\cite{imry.ma:1975}.

Most of our analysis is based on the distribution $P_{L,T}^{(k)}(\epsilon,s)$, 
defined as the probability to observe energy density $\epsilon$ and nematic 
order parameter $s$, at temperature $T$, system size $L$, and for some sample of 
random fields~$k$. We measure the distributions for $L=7-15$. The nematic order 
parameter $s$ is defined as the maximum eigenvalue of the 3D orientational 
tensor. In a perfectly aligned nematic sample $s=1$, while an isotropic sample 
yields $s \to 0$ in the thermodynamic limit. We use broad histogram methods, 
namely Wang-Landau sampling \cite{wang.landau:2001} and successive umbrella 
sampling \cite{virnau.muller:2004}, to obtain $P_{L,T}^{(k)}(\epsilon,s)$. These 
methods ensure that the simulation performs a random walk in phase space. This 
is crucial because the SSS model is known to exhibit meta-stable states 
\cite{rotunno:097802, citeulike:7427535}, in which standard Monte Carlo 
simulations (sampling directly from the Boltzmann distribution) may \ahum{get 
stuck}. Since we expect self-averaging to be violated, it is crucial that the 
distributions be measured for $k=1 , \ldots ,M$ random field samples, where $M$ 
must be large. We use $M \sim 1000-2500$, based on the convergence of 
\ahum{running averages} of quantities of interest onto plateau values. We also 
measure correlation functions; the latter are obtained for $L=30$ using standard 
Boltzmann sampling.

%% BEGIN FIGURE
\begin{figure}
\begin{center}
\includegraphics[width=0.8\columnwidth]{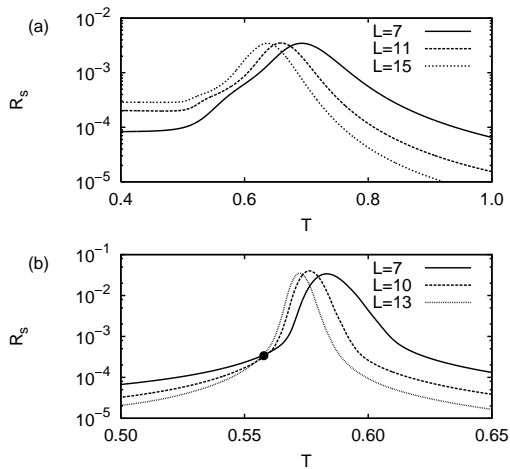}
\caption{\label{fig:fig1} $\rs$ versus $T$ using $p=2$ (a) and $p=10$ (b) 
for several $L$. The temperature where $\rs$ is maximal defines $T_R$. Note that 
$T_R$ decreases with $L$. For $p=2$, there is no self-averaging at low $T$. In 
contrast, for $p=10$, self-averaging is restored at low $T$, and a sharp phase 
transition occurs (marked with the dot).}
\end{center}
\end{figure}
%% END FIGURE

For each sample~$k$, we compute the thermally averaged nematic order parameter 
$\avg{s}_k$ and measure the fluctuation between samples $\rs^2 = [\avg{s}^2]- 
[\avg{s}]^2$, with $[\cdot]$ the disorder average $[X^n] = (1/M) \sum_{k=1}^M 
X_k^n$. If the system self-averages, $\rs \to 0$ in the thermodynamic limit, in 
which case a single experiment on a large system will be representative for all 
samples. In \fig{fig:fig1}(a), we plot $\rs$ versus $T$ for three system sizes. 
The striking result is that, at low temperature, $\rs$ does {\it not} decay to 
zero with increasing $L$ but remains finite. The onset to the non-self-averaging 
regime is marked by a maximum in $\rs$, at temperature $T=T_R$. We thus identify 
two regimes: a high-$T$ regime ($T>T_R$) where the SSS model self-averages 
($\rs$ decreases with $L$), and a low-$T$ regime ($T<T_R$) where self-averaging 
is violated ($\rs$ remains finite).

%% BEGIN FIGURE
\begin{figure}
\begin{center}
\includegraphics[width=0.8\columnwidth]{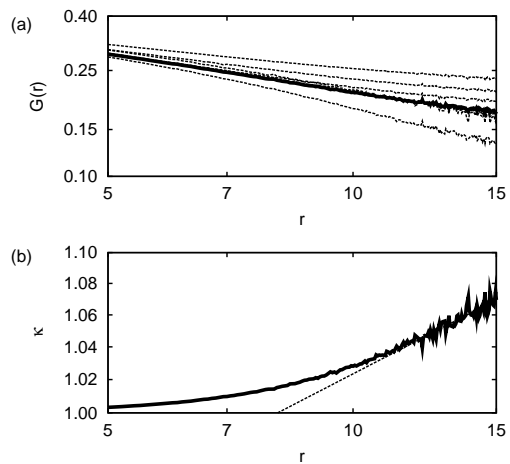}

\caption{\label{fig2} Correlation functions for $p=2$, $L=30$, and $T=0.5$ 
(which is well below $T_R$) on double logarithmic scales; due to periodic 
boundaries up to $r_{\rm max}=15$ can be sampled. (a) $G(r)$ obtained for 
several samples (dashed curves) together with the disorder-averaged result 
$[G(r)]$ (solid curve). (b) $\kappa$ versus $r$; the dashed line is a power law 
fit to the large $r$ regime.}

\end{center}
\end{figure}
%% END FIGURE

The violation of self-averaging at low $T$ profoundly affects the nematic 
correlation function $G(r) = \avg{ \frac{3}{2} (\vec{d}(0) \cdot \vec{d}(r))^2 - 
\frac{1}{2}}$ \cite{citeulike:5717604}. (In this work, $G(r)$ is calculated 
using all spins, i.e.~free and static ones.) Since it holds that 
$G(L/2)=\avg{s}^2$, with $L$ the edge of the simulation box, and since $\rs>0$, 
fluctuations in $G(r)$ between disorder samples are automatically implied. We 
must therefore consider $G_k(r)$, i.e.~the nematic correlation function obtained 
in the $k$-th random field sample. In the high-$T$ regime, we find that $G_k(r)$ 
decays exponentially to zero, with negligible fluctuations between samples: the 
SSS model is isotropic and self-averaging when $T>T_R$. In contrast, in the 
low-$T$ regime, $G_k(r)$ fluctuates profoundly between disorder samples 
(\fig{fig2}(a)). Note that we concentrate on the tail of $G(r)$ and so the range 
$r<5$ is discarded. In some samples, $G_k(r)$ decays very rapidly, while in 
others the decay is much slower. Clearly, when $T<T_R$ a single measurement of 
$G_k(r)$ is not representative.

The key point is that, in random field systems, there exist two correlation 
functions: the {\it connected} correlation function $[G(r)]$ (i.e.~the nematic 
correlation function averaged over all samples), and the {\it disconnected} 
correlation function $[G(r)^2]$ \cite{nattermann:1998, citeulike:5967178}. The 
solid curve in \fig{fig2}(a) shows $[G(r)]$: its decay to zero is most 
consistent with a power law, suggesting quasi-long-range order. This agrees with 
\olcite{citeulike:5967865}, but it disagrees with \olcite{citeulike:5717604} 
(where short-ranged exponential decay is observed). Regardless of the precise 
form of the decay, we confirm that $G_k(r)$ does not self-average. This is shown 
in \fig{fig2}(b), where $\kappa \equiv [G(r)^2]/[G(r)]^2$ is plotted. At large 
$r$, power law growth $\kappa \propto r^\theta$, with $\theta \sim 0.1$, is 
observed. The disconnected correlations thus decay independently from the 
connected ones. In contrast, if $G_k(r)$ were self-averaging, the fluctuation 
$[G(r)^2] - [G(r)]^2$ would be zero at large $r$: $[G(r)^2]$ and $[G(r)]^2$ then 
decay with the same exponent. Since the correlation functions do not 
self-average, it follows that properties extracted from these functions do not 
self-average either, which includes the correlation length $\xi$ 
\cite{citeulike:7465852, citeulike:3201143}. The Brout argument 
\cite{citeulike:7522088}, which conceives the thermodynamic limit as a large 
number of independent sub-samples of size $\xi$, thus breaks down. Instead, 
$\xi$ must be regarded as a random variable. The power law decay of $[G(r)]$ 
observed by us indicates that $\xi$ itself is very large, if not infinite. For 
$\xi \to \infty$, the Brout argument breaks down in any case 
\cite{citeulike:6348195}.

%% BEGIN FIGURE
\begin{figure}
\begin{center}
\includegraphics[width=0.85\columnwidth]{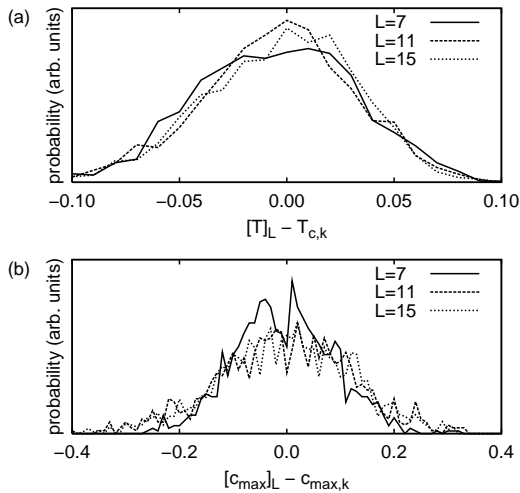}
\caption{\label{fig:fig3} Histograms of $T_{c,k}$ (a) and $c_{{\rm max},k}$ 
(b), shifted by their respective averages, and for several $L$. The 
histograms do not become sharp as $L$ increases.}
\end{center}
\end{figure}
%% END FIGURE

How does this affect the IN transition in the SSS model? The usual approach to 
detect the IN transition is to measure the specific heat $c=V(\avg{\epsilon^2} - 
\avg{\epsilon}^2)$ versus $T$; at the transition, $c$ reaches a maximum. For 
each random field sample~$k$, we measured the temperature $T_{c,k}$ where $c$ 
was maximal, and the corresponding value $c_{{\rm max},k}$. Since $\xi$ does not 
self-average, an unusually large fluctuation $[T_c^2] - [T_c]^2$ is expected. 
This is confirmed in \fig{fig:fig3}(a), where histograms of $T_{c,k}$ are shown, 
shifted such that $[T_c]$ is at zero, and for several $L$. The striking result 
is that the distributions do not become sharp as $L$ increases. The specific 
heat itself is also non-self-averaging. This is illustrated in 
\fig{fig:fig3}(b), where histograms of $c_{{\rm max},k}$ are shown, shifted by 
$[c_{\rm max}]$, and again for several $L$. We also observed that $[T_c]$ is 
very close to the temperature $T_R$ where $\rs$ is maximal. A signature of the 
onset to the low-$T$ regime (where self-averaging is violated) is thus also 
provided by the specific heat maximum. Both $[T_c]$ and $T_R$ decrease with 
increasing $L$: the non-self-averaging regime $T<T_R$ thus gets smaller in 
larger systems. Unfortunately, finite size scaling with a non-self-averaging 
correlation length is complicated -- a rigorous scaling theory remains elusive 
-- and so it is difficult to estimate $T_R$ in the thermodynamic limit. The 
decrease of $T_R$ with $L$, and hence of $[T_c]$, is in any case slow. For 
instance, if we assume a power law shift $T_R - T_\infty \propto 1/L^y$, 
$T_\infty \equiv \lim_{L \to \infty} T_R$, a fit to our data yields a maximum 
value for the exponent $y_{\rm max} \sim 0.16$; this upper bound is obtained by 
assuming $T_\infty=0$. 

To conclude: the SSS model (with $p=2$ in \eq{ll} and quenched spin fraction 
$q=0.1$) does not feature a sharp phase transition. For a given sample~$k$ of 
random fields, a well-defined temperature $T_{c,k}$ where the specific heat 
attains its maximum can be measured, but the fluctuation in $T_{c,k}$ between 
samples remains finite, even as $L \to \infty$. We attribute this behavior to 
the existence of a non-trivial {\it disconnected} correlation function, which 
implies a non-self-averaging correlation length when $T < T_R \sim [T_c]$. In 
this regime, the SSS model does not self-average. The temperature $T_R$ decays 
extremely slowly with system size; whether $T_R$ remains finite in the 
thermodynamic limit, or whether it decays to zero, cannot be discerned from our 
data. Since the decay of $T_R$ and $[T_c]$ with $L$ is slow, it is likely that 
the non-self-averaging regime survives in macroscopic samples (even if 
$T_\infty=0$). We expect that by varying $T$ a maximum in the specific heat will 
be found, but the value of the specific heat at the maximum will vary between 
samples. There is some experimental evidence for this behavior. The liquid 
crystal 8CB in bulk undergoes a weak first-order IN transition 
\cite{citeulike:5091293, citeulike:5930299}, as does \eq{ll} with $p=2$. Upon 
insertion in aerogel, the enthalpy obtained in different samples ranges from 
$3.6 - 5.23 \, \rm J/g$, which is unusually large \cite{citeulike:5091293}. 
However, since the enthalpy is related to the specific heat, and since the 
specific heat does not self-average (\fig{fig:fig3}(b)), a large enthalpy 
fluctuation between samples would, in fact, not be unexpected.

%% BEGIN FIGURE
\begin{figure}
\begin{center}
\includegraphics[width=0.8\columnwidth]{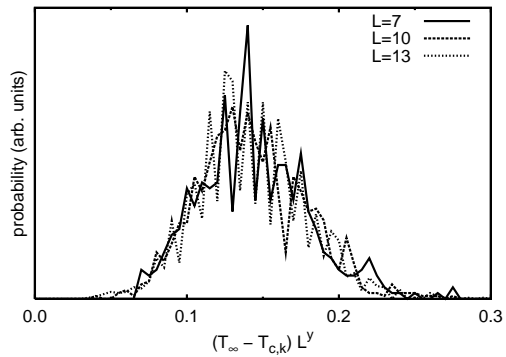}
\caption{\label{fig:fig4} Histograms of $(T_{\rm \infty} - T_{c,k}) L^y$ for 
$p=10$. The curves for different $L$ collapse, consistent with \eq{fss}.}
\end{center}
\end{figure}
%% END FIGURE

Do our results imply the absence of IN transitions, in general, in the presence 
of random field disorder? The answer to this question is an unequivocal 
\ahum{No}! The phase behavior of liquid crystals is not dictated by any 
universality class, and by changing details in the particle interaction 
qualitatively different scenarios may develop \cite{citeulike:7528527}. To 
illustrate this, we reconsider \eq{ll} using $p=10$; this makes the bulk IN 
transition more strongly first-order \cite{citeulike:7528527}. Again using a 
fraction of quenched spins $q=0.1$, we show in \fig{fig:fig1}(b) the variation 
of $\rs$ with $T$. The striking difference with $p=2$ is that self-averaging is 
restored at low temperatures. There now appears an intermediate regime of 
temperatures where self averaging is violated, but this regime becomes smaller 
as $L$ increases. Hence, in the thermodynamic limit, self-averaging is violated 
at only one temperature, which then reflects a sharp phase transition, with 
finite size effects given by \cite{aharony.harris:1996}
\begin{equation}\label{fss}
 \sqrt{[T_c^2] - [T_c]^2} \propto [T_c] - T_\infty \propto 1/L^y,
\end{equation}
where $T_\infty$ is the transition temperature in the thermodynamic limit. This 
implies that histograms of $(T_\infty - T_{c,k}) L^y$ become $L$-independent, 
provided correct values of $T_\infty$ and $y$ are used. The scaling is confirmed 
in \fig{fig:fig4}, using $T_\infty \approx 0.558$ and $y \approx 0.88$, and the 
collapse is clearly excellent. Incidentally, $T_\infty$ corresponds to an 
approximate intersection in curves of $\rs$ versus $T$ for different $L$ 
(\fig{fig:fig1}(b)), which offers an alternative route to locate the transition.

In summary: we have shown that the IN transition in the presence of random 
fields is strongly affected by a lack of self-averaging. Certainly for computer 
simulations, taking a disorder average $[\cdot]$ involving many samples is 
crucial. For a bulk IN transition that is weakly first-order, the violation of 
self-averaging is so severe, even the correlation length $\xi$ becomes 
non-self-averaging \cite{citeulike:7465852, citeulike:3201143}. This manifests 
itself from the nematic correlation function, which becomes strongly sample 
dependent. A consequence is that no sharp IN transition remains in this case. 
For a bulk IN transition that is more strongly first-order, the violation of 
self-averaging is restricted to a single temperature in the thermodynamic limit. 
In this case, a phase transition does occur, and finite size effects near the 
transition are well understood \cite{aharony.harris:1996}. As far as we know, a 
scaling theory for the case where $\xi$ does not self-average remains elusive. 
In some sense, the finite size effects observed by us for $p=2$ resemble those 
of \eq{fss}, but in the limit where $y \to 0$. Perhaps a new scaling theory 
should be developed keeping this in mind. 

\noindent This work was supported by the {\it Deutsche Forschungsgemeinschaft}, 
Emmy Noether program:~VI~483/1-1.

\vspace{-5mm}

\bibliography{refs}

\begin{thebibliography}{10}

\bibitem{citeulike:6711617}
Crawford GP and Z̆umer S (Eds.), Liquid Crystals in Complex Geometries (Taylor
  \& Francis, 1996)

\bibitem{citeulike:5091293}
Wu L, Zhou B, Garland CW, Bellini T, and Schaefer DW, \textit{Heat-capacity
  study of nematic-isotropic and nematic--smectic-A transitions for
  octylcyanobiphenyl in silica aerogels}, Phys.~Rev. E \textbf{51}, 2157 (1995)

\bibitem{citeulike:6721301}
Maritan A, Cieplak M, and Banavar JR, \textit{Nematic-Isotropic Transition in
  Porous Media}, in: GP~Crawford and S~\v{Z}umer (Eds.), Liquid Crystals in
  Complex Geometries, 483 (Taylor \& Francis, 1996)

\bibitem{citeulike:5930299}
Bellini T, Clark NA, Muzny CD, Wu L, Garland CW, Schaefer DW, and Oliver BJ,
  \textit{Phase behavior of the liquid crystal 8CB in a silica aerogel},
  Phys.~Rev. Lett. \textbf{69}, 788 (1992)

\bibitem{citeulike:6642489}
Wu Xl, Goldburg WI, Liu MX, and Xue JZ, \textit{Slow Dynamics of
  Isotropic-Nematic Phase Transition in Silica Gels}, Phys.~Rev. Lett.
  \textbf{69}, 470 (1992)

\bibitem{citeulike:6645610}
Iannacchione G and Finotello D, \textit{Confinement and orientational study at
  liquid crystal phase transitions}, Liq.~Cryst. \textbf{14}, 1135 (1993)

\bibitem{citeulike:6642471}
Tripathi S, Rosenblatt C, and Aliev FM, \textit{Orientational susceptibility in
  porous glass near a bulk nematic-isotropic phase transition}, Phys.~Rev.
  Lett. \textbf{72}, 2725 (1994)

\bibitem{citeulike:6642981}
Zhou B, Iannacchione GS, Garland CW, and Bellini T, \textit{Random-field
  effects on the nematic--smectic-A phase transition due to silica aerosil
  particles}, Phys.~Rev. E \textbf{55}, 2962 (1997)

\bibitem{imry.ma:1975}
Imry Y and Ma SK, \textit{Random-Field Instability of the Ordered State of
  Continuous Symmetry}, Phys. Rev. Lett. \textbf{35}, 1399 (1975)

\bibitem{citeulike:5717604}
Bellini T, Buscaglia M, Chiccoli C, Mantegazza F, Pasini P, and Zannoni C,
  \textit{Nematics with Quenched Disorder: What Is Left when Long Range Order
  Is Disrupted?}, Phys.~Rev. Lett. \textbf{85}, 1008 (2000)

\bibitem{citeulike:6645961}
Cleaver DJ, Kralj S, Sluckin TJ, and Allen MP, \textit{The Random Anisotropy
  Nematic Spin Model}, in: GP~Crawford and S~\v{Z}umer (Eds.), Liquid Crystals
  in Complex Geometries, 467 (Taylor \& Francis, 1996)

\bibitem{citeulike:6657736}
Feldman DE and Pelcovits RA, \textit{Liquid crystals in random porous media:
  Disorder is stronger in low-density aerosils}, Phys.~Rev. E \textbf{70},
  040702 (2004)

\bibitem{citeulike:5967865}
Feldman DE, \textit{Quasi-Long-Range Order in Nematics Confined in Random
  Porous Media}, Phys.~Rev. Lett. \textbf{84}, 4886 (2000)

\bibitem{citeulike:6721317}
Clark NA, Bellini T, Malzbender RM, Thomas BN, Rappaport AG, Muzny CD, Schaefer
  DW, and Hrubesh L, \textit{X-ray scattering study of smectic ordering in a
  silica aerogel}, Phys.~Rev. Lett. \textbf{71}, 3505 (1993)

\bibitem{citeulike:4197313}
Imry Y and Wortis M, \textit{Influence of quenched impurities on first-order
  phase transitions}, Phys.~Rev. B \textbf{19}, 3580 (1979)

\bibitem{citeulike:6642433}
Iannacchione GS, Crawford GP, \v{Z}umer S, Doane JW, and Finotello D,
  \textit{Randomly constrained orientational order in porous glass}, Phys.~Rev.
  Lett. \textbf{71}, 2595 (1993)

\bibitem{citeulike:6645630}
Kralj S, Lahajnar G, Zidan\v{s}ek A, Kopa\v{c} NV, Vilfan M, Blinc R, and Kosec
  M, \textit{Deuterium NMR of a pentylcyanobiphenyl liquid crystal confined in
  a silica aerogel matrix}, Phys.~Rev. E \textbf{48}, 340 (1993)

\bibitem{citeulike:6657729}
Mercuri F, Paoloni S, Zammit U, and Marinelli M, \textit{Dynamics at the
  Nematic-Isotropic Phase Transition in Aerosil Dispersed Liquid Crystal},
  Phys.~Rev. Lett. \textbf{94}, 247801 (2005)

\bibitem{citeulike:6642390}
Bellini T, Clark NA, and Schaefer DW, \textit{Dynamic Light Scattering Study of
  Nematic and Smectic- $A$ Liquid Crystal Ordering in Silica Aerogel},
  Phys.~Rev. Lett. \textbf{74}, 2740 (1995)

\bibitem{rotunno:097802}
Rotunno M, Buscaglia M, Chiccoli C, Mantegazza F, Pasini P, Bellini T, and
  Zannoni C, \textit{Nematics with Quenched Disorder: Pinning out the Origin of
  Memory}, Phys.~Rev. Lett. \textbf{94}, 097802 (2005)

\bibitem{citeulike:5887661}
Petridis L and Terentjev EM, \textit{Nematic-isotropic transition with quenched
  disorder}, Phys.~Rev. E \textbf{74}, 051707 (2006)

\bibitem{aharony.harris:1996}
Aharony A and Harris AB, \textit{Absence of Self-Averaging and Universal
  Fluctuations in Random Systems near Critical Points}, Phys.~Rev. Lett.
  \textbf{77}, 3700 (1996)

\bibitem{citeulike:6672115}
Wiseman S and Domany E, \textit{Finite-Size Scaling and Lack of Self-Averaging
  in Critical Disordered Systems}, Phys.~Rev. Lett. \textbf{81}, 22 (1998)

\bibitem{citeulike:6348195}
Wiseman S and Domany E, \textit{Self-averaging, distribution of pseudocritical
  temperatures, and finite size scaling in critical disordered systems},
  Phys.~Rev. E \textbf{58}, 2938 (1998)

\bibitem{citeulike:7465852}
Parisi G, Picco M, and Sourlas N, \textit{Scale invariance and self-averaging
  in disordered systems}, Europhys. Lett. \textbf{66}, 465 (2004)

\bibitem{citeulike:3201143}
Parisi G and Sourlas N, \textit{Scale Invariance in Disordered Systems: The
  Example of the Random-Field Ising Model}, Phys.~Rev. Lett. \textbf{89},
  257204 (2002)

\bibitem{citeulike:6711320}
Bellini T, Chiccoli C, Pasini P, and Zannoni C, \textit{Lattice Spin Models of
  Liquid Crystals in Aerogels}, Mol.~Cryst.~Liq.~Cryst. \textbf{290}, 227
  (1996)

\bibitem{citeulike:5091220}
Chakrabarti J, \textit{Simulation Evidence of Critical Behavior of
  Isotropic-Nematic Phase Transition in a Porous Medium}, Phys.~Rev. Lett.
  \textbf{81}, 385 (1998)

\bibitem{citeulike:5091176}
Maritan A, Cieplak M, Bellini T, and Banavar JR, \textit{Nematic-Isotropic
  Transition in Porous Media}, Phys.~Rev. Lett. \textbf{72}, 4113 (1994)

\bibitem{physreva.6.426}
Lebwohl PA and Lasher G, \textit{Nematic-Liquid-Crystal Order: A Monte Carlo
  Calculation}, Phys.~Rev. A \textbf{6}, 426 (1972)

\bibitem{wang.landau:2001}
Wang F and Landau DP, \textit{Efficient, Multiple-Range Random Walk Algorithm
  to Calculate the Density of States}, Phys.~Rev. Lett. \textbf{86}, 2050
  (2001)

\bibitem{virnau.muller:2004}
Virnau P and M\"{u}ller M, \textit{Calculation of free energy through
  successive umbrella sampling}, J.~Chem.~Phys. \textbf{120}, 10925 (2004)

\bibitem{citeulike:7427535}
Buscaglia M, Bellini T, Chiccoli C, Mantegazza F, Pasini P, Rotunno M, and
  Zannoni C, \textit{Memory effects in nematics with quenched disorder},
  Phys.~Rev. E \textbf{74}, 011706 (2006)

\bibitem{nattermann:1998}
Nattermann T, \textit{Theory of the Random Field Ising Model}, in: AP~Young
  (Ed.), Spin Glasses and Random Fields, 277 (World Scientific, Singapore,
  1998)

\bibitem{citeulike:5967178}
Khasanov B, \textit{Isotropic phase of nematics in porous media}, JETP Lett.
  \textbf{81}, 24 (2005)

\bibitem{citeulike:7522088}
Brout R, \textit{Statistical Mechanical Theory of a Random Ferromagnetic
  System}, Phys.~Rev. \textbf{115}, 824 (1959)

\bibitem{citeulike:7528527}
Fish JM and Vink RLC, \textit{Isotropic-to-nematic transition in confined
  liquid crystals: An essentially nonuniversal phenomenon}, Phys.~Rev. E
  \textbf{81}, 021705 (2010)

\end{thebibliography}

\end{document}